\RequirePackage{fix-cm}
\documentclass[twocolumn,epjc3]{svjour3}  
\smartqed  
\RequirePackage{graphicx}
\usepackage{lineno}
\begin{document}
\title{Search for Low-Energy Signals from Fast Radio Bursts with the Borexino Detector}





\author{
S.~Appel\thanksref{Munchen}
\and
Z.~Bagdasarian\thanksref{Berkeley,Juelich}
\and
D.~Basilico\thanksref{Milano}
\and
G.~Bellini\thanksref{Milano}
\and
J.~Benziger\thanksref{PrincetonChemEng}
\and
R.~Biondi\thanksref{LNGS}
\and
B.~Caccianiga\thanksref{Milano}
\and
F.~Calaprice\thanksref{Princeton}
\and
A.~Caminata\thanksref{Genova}
\and
A.~Chepurnov\thanksref{Lomonosov}
\and
D.~D'Angelo\thanksref{Milano}
\and
A.~Derbin\thanksref{Peters,Kurchatov}
\and
A.~Di Giacinto\thanksref{LNGS}
\and
V.~Di Marcello\thanksref{LNGS}
\and
X.F.~Ding\thanksref{Princeton}
\and
A.~Di Ludovico\thanksref{Princeton} 
\and
L.~Di Noto\thanksref{Genova}
\and
I.~Drachnev\thanksref{Peters}
\and
D.~Franco\thanksref{APC}
\and
C.~Galbiati\thanksref{Princeton,GSSI}
\and
C.~Ghiano\thanksref{LNGS}
\and
M.~Giammarchi\thanksref{Milano}
\and
A.~Goretti\thanksref{LNGSG,Princeton} 
\and
A.S.~G\"ottel\thanksref{Juelich,RWTH}
\and
M.~Gromov\thanksref{Lomonosov,Dubna}
\and
D.~Guffanti\thanksref{Milanoa,Mainz}
\and
Aldo~Ianni\thanksref{LNGS}
\and
Andrea~Ianni\thanksref{Princeton}
\and
A.~Jany\thanksref{Krakow}
\and
V.~Kobychev\thanksref{Kiev}
\and
G.~Korga\thanksref{London,Atomki}
\and
S.~Kumaran\thanksref{Juelich,RWTH}
\and
M.~Laubenstein\thanksref{LNGS}
\and
E.~Litvinovich\thanksref{Kurchatov,Kurchatovb}
\and
P.~Lombardi\thanksref{Milano}
\and
I.~Lomskaya\thanksref{Peters}
\and
L.~Ludhova\thanksref{Juelich,RWTH}
\and
G.~Lukyanchenko\thanksref{Kurchatov}
\and
I.~Machulin\thanksref{Kurchatov,Kurchatovb}
\and
J.~Martyn\thanksref{Mainz}
\and
E.~Meroni\thanksref{Milano}
\and
L.~Miramonti\thanksref{Milano}
\and
M.~Misiaszek\thanksref{Krakow}
\and
V.~Muratova\thanksref{Peters}
\and
R.~Nugmanov\thanksref{Kurchatov,Kurchatovb}
\and
L.~Oberauer\thanksref{Munchen}
\and
V.~Orekhov\thanksref{Mainz}
\and
F.~Ortica\thanksref{Perugia}
\and
M.~Pallavicini\thanksref{Genova}
\and
L.~Pelicci\thanksref{Juelich,RWTH}
\and
\"O.~Penek\thanksref{Juelich}
\and
L.~Pietrofaccia\thanksref{Princeton}
\and
N.~Pilipenko\thanksref{Peters}
\and
A.~Pocar\thanksref{UMass}
\and
G.~Raikov\thanksref{Kurchatov}
\and
M.T.~Ranalli\thanksref{LNGS}
\and
G.~Ranucci\thanksref{Milano}
\and
A.~Razeto\thanksref{LNGS}
\and
A.~Re\thanksref{Milano}
\and
M.~Redchuk\thanksref{Padova,Juelich,RWTH} 
\and
N.~Rossi\thanksref{LNGS}
\and
S.~Sch\"onert\thanksref{Munchen}
\and
D.~Semenov\thanksref{Peters}
\and
G.~Settanta\thanksref{Romaa,Juelich}
\and
M.~Skorokhvatov\thanksref{Kurchatov,Kurchatovb}
\and
A.~Singhal\thanksref{Juelich,RWTH}
\and
O.~Smirnov\thanksref{Dubna}
\and
A.~Sotnikov\thanksref{Dubna}
\and
R.~Tartaglia\thanksref{LNGS}
\and
G.~Testera\thanksref{Genova}
\and
E.~Unzhakov\thanksref{Peters}
\and
A.~Vishneva\thanksref{Dubna}
\and
R.B.~Vogelaar\thanksref{Virginia}
\and
F.~von~Feilitzsch\thanksref{Munchen}
\and
M.~Wojcik\thanksref{Krakow}
\and
M.~Wurm\thanksref{Mainz}
\and
S.~Zavatarelli\thanksref{Genova}
\and
I.~Zhutikov\thanksref{Kurchatov,Kurchatovb}
\and
K.~Zuber\thanksref{Dresda}
\and
G.~Zuzel\thanksref{Krakow}
}

\thankstext{Berkeley}{Present address: University of California, Berkeley, Department of Physics, CA 94720, Berkeley, USA}
\thankstext{LNGSG}{Present address: INFN Laboratori Nazionali del Gran Sasso, 67010 Assergi (AQ), Italy}
\thankstext{Milanoa}{Present address: Dipartimento di Fisica, Universit\'a degli Studi e INFN Milano-Bicocca, 20126 Milano, Italy}
\thankstext{Padova}{Present address: Dipartimento di Fisica e Astronomia dell' Universit\'a di Padova and INFN Sezione di Padova, Padova, Italy}
\thankstext{Romaa}{Present address: Istituto Superiore per la Protezione e la Ricerca Ambientale, 00144 Roma, Italy}

\institute{\bf{The Borexino Collaboration}}
\institute{Physik-Department, Technische Universit\"at  M\"unchen, 85748 Garching, Germany\label{Munchen}
\and
Institut f\"ur Kernphysik, Forschungszentrum J\"ulich, 52425 J\"ulich, Germany\label{Juelich}
\and
Dipartimento di Fisica, Universit\`a degli Studi e INFN, 20133 Milano, Italy\label{Milano}
\and
Chemical Engineering Department, Princeton University, Princeton, NJ 08544, USA\label{PrincetonChemEng}
\and
INFN Laboratori Nazionali del Gran Sasso, 67010 Assergi (AQ), Italy\label{LNGS}
\and
Physics Department, Princeton University, Princeton, NJ 08544, USA\label{Princeton}
\and
Dipartimento di Fisica, Universit\`a degli Studi e INFN, 16146 Genova, Italy\label{Genova}
\and
Lomonosov Moscow State University Skobeltsyn Institute of Nuclear Physics, 119234 Moscow, Russia\label{Lomonosov}
\and
St. Petersburg Nuclear Physics Institute NRC Kurchatov Institute, 188350 Gatchina, Russia\label{Peters}
\and
National Research Centre Kurchatov Institute, 123182 Moscow, Russia\label{Kurchatov}
\and
APC, Universit\'e de Paris, CNRS, Astroparticule et Cosmologie, Paris F-75013, France\label{APC}
\and
Gran Sasso Science Institute, 67100 L'Aquila, Italy\label{GSSI}
\and
Joint Institute for Nuclear Research, 141980 Dubna, Russia\label{Dubna}
\and
Institute of Physics and Excellence Cluster PRISMA+, Johannes Gutenberg-Universit\"at Mainz, 55099 Mainz, Germany\label{Mainz}
\and
M.~Smoluchowski Institute of Physics, Jagiellonian University, 30348 Krakow, Poland\label{Krakow}
\and
Institute for Nuclear Research of NAS Ukraine, 03028 Kyiv, Ukraine\label{Kiev}
\and
RWTH Aachen University, 52062 Aachen, Germany\label{RWTH}
\and
National Research Nuclear University MEPhI (Moscow Engineering Physics Institute), 115409 Moscow, Russia\label{Kurchatovb}
\and
Dipartimento di Chimica, Biologia e Biotecnologie, Universit\`a degli Studi e INFN, 06123 Perugia, Italy\label{Perugia}
\and
Amherst Center for Fundamental Interactions and Physics Department, UMass, Amherst, MA 01003, USA\label{UMass}
\and
Department of Physics, Technische Universit\"at Dresden, 01062 Dresden, Germany\label{Dresda}
\and
Physics Department, Virginia Polytechnic Institute and State University, Blacksburg, VA 24061, USA\label{Virginia}
\and
Department of Physics, Royal Holloway, University of London, Department of Physics, School of Engineering, Physical and Mathematical Sciences, Egham, Surrey, TW20 OEX, UK\label{London}
\and
Institute of Nuclear Research (Atomki), Debrecen, Hungary\label{Atomki}
}
\date{Received: date / Accepted: date}

\maketitle

\begin{abstract}
The search for neutrino events in correlation with 42 most intense fast radio bursts (FRBs) has been performed using the Borexino dataset from 05/2007 to 06/2021.
We have searched for signals with visible energies above $250$~keV within a time window of $\pm1000$~s corresponding to detection time of a particular FRB.
We also applied an alternative approach based on searching for specific shapes of neutrino-electron scattering spectra in the full exposure data of the Borexino detector.
In particular, two incoming neutrino spectra were considered: the monoenergetic line and the spectrum expected from supernovae.
The same spectra were considered for electron antineutrinos detected through inverse beta-decay reaction.
No statistically significant excess over the background was observed.
As a result, the strongest upper limits on FRB-associated neutrino fluences of all flavors have been obtained in the $0.5 - 50$~MeV neutrino energy range.

\keywords{Fast radio bursts \and neutrino \and Borexino}
\end{abstract}

\section{Introduction}\label{intro}
A fast radio burst (FRB) is a millisecond radio transient observed at extragalactic or cosmological distance.
Although FRBs were discovered almost 15 years ago~\cite{Lor2007}, the nature of their source remains unclear.
Numerous models with a wide variety of physical processes have been proposed to explain the origin of FRBs (see review articles~\cite{Pop2018,Pet2019,Cor2019,Pla2020}).
The most popular class of models postulates a production mechanism associated with arising activity of magnetars~\cite{Pop2013,Lyu2014}.
These models have received support from repeating behavior of some FRBs, especially after detection of FRB200428 in temporal and spatial coincidence with X-ray burst from magnetar SGR~1935+2154 in the Milky Way galaxy~\cite{Boc2020,And2020}. 

Single-burst FRB models involve processes of supernova evolution, mergers, and collapses of neutron stars~\cite{Rav2014,Fal2014,Tot2013,Wan2016} with emission of neutrinos~\cite{Met2008,Fah2017,Xia2014,Gup2018} (and possibly axions) which could be potentially detected by large-volume Cherenkov or scintillation detectors.
The IceCube Neutrino Observatory has searched for spatial and temporal correlation between events with energies above $50$~GeV, as well as temporal correlation between MeV events and 28 FRBs.
It has set  the upper limits on neutrino fluences associated with them~\cite{Khe2017,Aar2018,Aar2020}.
The ANTARES Neutrino Telescope looked for TeV-PeV high-energy neutrinos spatially and temporally coincident with FRBs detected during 2013-2017, but no coincident neutrino candidate was observed~\cite{Alb2019}.

Borexino, a real-time liquid scintillator detector designed for solar neutrino spectroscopy, is located at the Gran Sasso National Laboratory, in Italy~\cite{Ali2002,Ali2009,Bel2014,Ago2019}.
Due to its extremely low background level, large target mass, and low energy threshold, the Borexino detector has been successfully used for studying low-energy neutrino fluxes from such transients as $\gamma$-ray bursts (GRBs), gravitational wave (GW) events and solar flares~\cite{Bel2011,Ago2017a,Ago2017b,Ago2021}.

Since modern radio telescopes have a narrow field of view, they are able to register only a fraction of the overall number of occurring FRBs.
Total expected all-sky event rate of FRBs with fluence above $\sim 2$~Jy$~$ms is roughly $2\times10^3$ per day~\cite{Pet2019,Bha2018}\footnote{Jy (Jansky) is a non-SI unit of spectral flux density, $1\ \rm{ Jy} = 10^{-26}$~W$~$m$^{-2}~$Hz$^{-1}$}.
Scintillation detectors lack the directional sensitivity to  incoming neutrinos and therefore can not be used for temporal analysis in a wide time window due to the high rate of FRBs (in contrast to GRBs or GW events) and a possible delay of low-energy neutrino signal arriving from extragalactic distances. 

Here, we have performed the temporal correlation analysis between Borexino events with visible energies above $0.25$~MeV and some of the most intensive FRBs assuming the direct connection between radio and neutrino fluences of an FRB.
Another approach was based on the search for the characteristic shape of ($\nu, e$)-scattering in the high-statistic Borexino spectrum.
Finally, we have taken into account that no events were observed with an energy greater than a certain value in the Borexino spectrum.
Two different spectra of incoming neutrinos ($\nu_{e,\mu,\tau}$ and $\bar{{\nu}}_{e,\mu,\tau}$) were used for the analysis: the monoenergetic line and the spectrum expected from supernovae.
The same $\bar{{\nu}}_{e}$-neutrino spectra were considered for detection with the inverse beta-decay reaction (IBD).

\section{Borexino Detector}
Borexino is a~real-time liquid scintillator detector for~solar neutrino spectroscopy. It is located underground at the Gran Sasso Laboratory (Italy) at a depth of $3400$~m.w.e.
Its~main goal is to~measure low-energy solar neutrinos via~($\nu,e$) scattering in an ultrapure liquid scintillator.

The inner vessel of the detector (IV) comprises $278$~tons of~purified organic liquid scintillator confined in a transparent nylon sphere of~$4.25$~m in radius. 
The scintillator was produced from petrochemical organics extracted from underground to fulfill high radiopurity requirements.
The scintillator compound is based on~pseudocumene (PC, $\rm{C_9H_{12}}$) doped with $1.5$~g/L of PPO ($\rm{C_{15}H_{11}NO}$).
It~is surrounded by~two concentric PC buffers ($323$ and $567$~tons, respectively) doped with a small amount of light quencher (dimethylphthalate, DMP) intended for light yield reduction.
The buffer partitioning is performed in order to reduce the diffusion of radon into the scintillator bulk.
The IV and the buffers are contained inside a~stainless steel sphere (SSS) with a diameter of $6.75$~m fixed in position by a~stainless steel support structure.
The SSS is enclosed in a~cylinder with a hemispheric top with a diameter~of $18$~m and height~of $16.9$~m.

The water tank (WT) is constructed of stainless steel with high radiopurity and contains $2100$~tons of ultrapure water as additional shielding against external $\gamma$-rays and neutrons.
The WT is equipped with~$208$~8-inch PMTs and serves as the Cherenkov muon veto (outer detector, OD) for identification of residual muons crossing the detector.
The scintillation light is detected by nominally~$2212$~8-inch PMTs of~the inner detector (ID) uniformly distributed on~the inner surface of~the~SSS.

Borexino detects charged particles via scintillation light produced in the liquid scintillator.
Each event occurring in~the detector is characterized by a~number of~fired PMTs whose pulse amplitudes and arrival times are recorded.
The number of active PMTs has slowly declined over the course of the Borexino data taking. 
For each FRB event considered in this analysis, the true number of active PMTs was included in the modeling of the detector response and the signal normalized appropriately.
These data are used to reconstruct the energy and spatial coordinates of the event and to identify the type of the particle $(e,\alpha,\mu)$. 
Both energy and spatial resolutions of the detector were studied with radioactive sources placed at different positions inside the inner vessel \cite{Bac2012}.
The energy and position resolutions are $\sigma_E \approx 50$~keV and $\sigma_X \approx 10$~cm at $1$~MeV with 2000 PMTs, respectively; both scaling with the energy of the event as $\sim 1/\sqrt{E}$ at low energies. 
The Borexino detector is unable to provide sufficient directional information about a single event due to the nearly isotropic emission of scintillation light (see, however, \cite{Ago2021e}). 

A more detailed description of the Borexino detector can be found in the following papers~\cite{Ali2002,Ali2009,Bel2014,Ago2019}.

Neutrinos ($\nu_x, x=e,\mu,\tau$) and antineutrinos ($\bar{\nu_x}$) are detected by means of their elastic scattering on electrons:
\begin{equation}\label{eq:1}
    \nu_x+e^- \rightarrow \nu_x+e^-\,, \quad \bar{\nu}_x+e^- \rightarrow \bar{\nu}_x+e^-\,.
\end{equation}
For a given neutrino energy, the maximum electron recoil energy is given by the Compton formula:
\begin{equation}\label{eq:2}
    E_{e_{max}}=2E_\nu^2/(2E_\nu+m_e),
\end{equation}
where $E_{\nu}$ is the (anti)neutrino energy and $m_e$ is the electron mass.
The interaction between the scattered electron and scintillator molecules produces photons which are registered by PMTs.

Electron antineutrinos $\bar{\nu}_e$ can also be detected via the inverse beta-decay (IBD) reaction with an energy threshold of $1.8$~MeV:
\begin{equation}\label{eq:3}
    \bar{\nu}_e+p \rightarrow n + e^+.
\end{equation}
The visible energy of the positron and two annihilation photons is related to the antineutrino energy as $E_{vis} = E_{\bar{\nu}_e} -~0.784~$MeV.
The neutron capture on protons produces a $2.22$~MeV $\gamma$-ray providing a delayed signal with the mean capture time of $\sim 260$~$\mu$s \cite{Bel2011d}.
In~contrast to the total cross section of $(\nu_e,e)$-scattering which is proportional to $E_\nu$ when $E_\nu \gg m_e$, the cross section of the IBD is proportional to $\sim E_{\bar {\nu}_e}^2$. 

Borexino was the first experiment to detect and then precisely measure the $^7\rm{Be}$ solar neutrino flux~\cite{Arp2008,Bel2011a} as well as the $\rm{^8B}$-neutrino flux with $3$~MeV threshold~\cite{Bel2010,Ago2020}. 
It also observed $pep$-neutrinos for the first time~\cite{Bel2012} and  made the first spectral measurement of $pp$-neutrinos~\cite{Bel2014a,Ago2018,Ago2019}, and provided the first experimental evidence  of solar neutrinos produced in the CNO cycle~\cite{Ago2020a,Ago2020e}.
The Borexino detector also registered antineutrinos $\bar{\nu}_e$ emitted in the decay of radionuclides naturally occurring in the Earth~\cite{Ago2015,Ago2020b,Bel2010g,Bel2013g}.

Due to its excellent radiopurity, large target mass and low energy threshold, Borexino is perfectly suited for the study of other fundamental problems, as well as searching for rare and exotic processes in particle physics and astrophysics.

The Borexino experiment obtained new data on solar neutrino properties: ruled out any significant day-night asymmetry of the 7Be neutrino interaction rate~\cite{Bel2012a}, set new limits on the effective magnetic moment of solar neutrinos~\cite{Arp2008,Ago2017}, on the flux of~$\bar{\nu}_e$ from the Sun~\cite{Bel2011c,Ago2021} and on the non-standard solar neutrino interactions ~\cite{Aga2020}. 
A search for a number of rare low-energy processes has been carried out: possible violation of the Pauli exclusion principle~\cite{Bel2010a}, high-energy solar axions~\cite{Bel2012b}, heavy sterile neutrino mixing in the $\rm{^8B}~\beta^+$-decay~\cite{Bel2013}, decay of an electron into a neutrino and a photon~\cite{Ago2015a}. 
Additionally, temporal correlations with transient astrophysical sources such as $\gamma$-ray bursts~\cite{Ago2017a}, gravitational wave events~\cite{Ago2017b}, and solar flares~\cite{Ago2021} have been performed.


\section{Data Selection}

Since a FRB is a very frequent event that is detected in a few cases, the analysis was conducted via two different approaches: a transient event search in a fixed time window with respect to the FRB detection moment and a generic search for extra neutrino-induced electron recoils or IBD components in the detector energy spectrum.

\subsection{Transient Approach Data Selection}
The first approach does not require a well-described spectral shape due to the limited number of events in the time window  and does not need any profound background reduction techniques, although the background is to be kept substantially low.
Neutrino--electron elastic scattering events which are of interest in the current analysis lack any characteristic interaction signature.
Thus, the background reduction has to be performed in a generic manner so as decrease in the detector count rate per unit of exposure.
The background of the Borexino detector includes the following main components:
\begin{itemize}
    \item Short-lived cosmogenic backgrounds ($\tau \leq 0.3$~s) produced inside the detector fiducial volume, such as $^{12}$B, $^{8}$He, $^{9}$C, $^{9}$Li etc.
    \item Other cosmogenic backgrounds produced inside the detector fiducial volume, such as longer-lived isotopes $^{11}$Be, $^{10}$C, $^{11}$C etc.
    \item External gamma-background associated with natural radioactivity in detector  materials and PMTs. 
    \item Backgrounds of the inner nylon vessel associated with radioactivity of the $^{210}$Pb and uranium/thorium decay chains.
    \item Natural backgrounds contained in the bulk of the detector fluid, such as $^{14}$C, $^{85}$Kr, $^{210}$Bi, and $^{210}$Pb.
    \item Solar neutrino recoil electrons of the $pp$-chain and the CNO cycle.
\end{itemize}  
These backgrounds can be suppressed by using information from the processed detector data, such as ID/OD coincidences as well as the energy and position reconstruction.
Cosmogenic backgrounds can be reduced by applying the detector time vetoes after each muon event that can be discriminated through coincidence with the outer veto as well as via pulse-shape discrimination~\cite{Bel2011d,Ago2021a}.
A veto length of $0.3$~s following a muon event is applied to suppress $^{12}$B to a statistically insignificant level and reduce $^{8}$He, $^{9}$C, and $^{9}$Li by a factor of 3 with a live time loss of $1$~\%.

Background components contained in the bulk can be reduced by applying a cut on the visible energy.
This is important specifically due to the presence of $^{14}$C in the scintillator.
The carbon isotope $^{14}$C produces a $\beta$-spectrum with an endpoint of $156$~keV and has an activity of roughly $110$~Bq in the whole inner vessel.
The presence of this spectral component sets the lower threshold of the analysis at~$250$~keV of the visible energy\footnote{The visible energy spectrum of $^{14}$C is broadened up to this value due to the detector energy resolution  and its pile-up}.

Background components contained in the nylon of the IV and other detector materials
cannot be removed by any kind of purification and therefore are $10^2 - 10^3$ times higher than within the bulk of the scintillator.
The most dangerous components are the $^{214}$Bi and $^{208}$Tl decays.
These nuclides undergo $\beta$ and $\beta+\gamma$ decay processes with continuous spectra overlapping with the energy region used in this analysis.
The only way to overcome this kind of background is to perform a geometrical cut on events, selecting those within the fiducial volume.
In our case the fiducial volume is defined such that all events within $75$~cm to the IV are discarded, which corresponds to 3 standard deviations of position reconstruction uncertainty at the lowest energy threshold.
The corresponding fiducial volume has a mass of $145$~t.
The Borexino dataset from May 15, 2007, (corresponding to the detector operation start) to June 21, 2021, was used for the temporal correlation analysis between the Borexino signals and the most intensive FRBs.

\subsection{Spectral Approach Data Selection}

The spectral approach to data selection is more complicated since background components should not only be sufficiently suppressed, but also well described in terms of spectral shapes.
Minimization of the number of these spectral components also benefits the final result due to suppression of their spectral correlations fit procedure, even if it comes at the expense of exposure loss.
Thus, we modify the data selection procedure of the transient analysis in the following way:
\begin{itemize}
    \item We use the most radiopure dataset acquired from January 01, 2013, (corresponding to the detector stabilization after the water extraction procedure finished at the end of 2011) to November 31, 2020.
    \item We apply an advanced system of the cosmogenic veto based on time and position reconstruction of muons and neutrons in the post-muon gate that strongly reduces cosmogenic backgrounds. 
\end{itemize}

The advanced cosmogenic veto system dedicated to the discrimination of short-lived cosmogenic nuclides is based on the information from the muon trigger and post-muon trigger of the Borexino detector. We organize it in a way, similar to identification of the cosmogenic $^{11}$C background performed in \cite{Ago2021a}, but with different times and radii, namely:

\begin{itemize}
    \item $120$~s full detector veto after each muon that crossed the ID and has more than 20 neutron-like daughters within the following $1.6$~ms trigger gate.
    \item $20$~s veto on the cylinder with a radius of $0.8$~m aligned with the muon track in case the track is reconstructed with OD signals.
    \item $120$~s spherical veto with a radius of $1.3$~m on each reconstructed neutron position in the $1.6$~ms post-muon trigger gate.
    \item full detector veto of $4$~s after each muon crossing the ID.
\end{itemize}
This veto system comes at the cost of $15.8$\% exposure loss that is calculated with toy Monte Carlo (or, in other words by numeric integration with Monte-Carlo method) with fake events produced uniformly within the ID with the constant rate of $100$~Hz.

The obtained spectrum is followed by the statistical subtraction of external background
 based on the radial distribution fit of each energy bin with the function described as:
\begin{equation}\label{eq:4}
    N(R) = N_0(R)\times(A+B\times exp(\lambda R ))\,,
\end{equation}
where $N_0(R)$ is the radial distribution of toy Monte Carlo events within the fiducial volume, $\lambda$ is a free parameter that describes external gamma-background attenuation, and A and B correspond to the internal and external signal, respectively.
Validity of this procedure was tested using the full detector Monte Carlo data~\cite{Ago2018a} and provides reliable separation of external backgrounds which comes at the cost of increased uncertainty in each bin.
The spectrum obtained corresponds to $298.4$~$\mathrm{kt}~\mathrm{day}$ of exposure and has a significantly reduced background composition that can be described by known background components.

\section{Analysis and Results}

As mentioned above, two different approaches have been used for studying the neutrino fluence associated with FRBs. 
First, the excess of the number of Borexino events was searched for in temporal correlation with the most intense FRBs selected from the existing databases.
Then, the Borexino energy spectrum with high statistics was analyzed in order to determine the possible additional unaccounted contribution from $(\nu,e)$-scattering and inverse beta-decay reactions.

\subsection{Selection of FRBs from Databases}

We used the \verb|chime-frb.ca| database accumulated by the CHIME Radio Telescope~\cite{Ami2018,Ami2021} and the \verb|frbcat.org| database that collected and summarized the data from several other telescopes, such as Parkes~\cite{Yan2021}, Arecibo~\cite{Arecibo}, Green Bank~\cite{GRBNK}, UTMOST~\cite{utmost1,utmost2}, ASKAP~\cite{Ban2017}, FAST~\cite{fst}, Apertif~\cite{apertif,apertif1}, VLA~\cite{vla}, DSA-19~\cite{dsa}, and Pushchino~\cite{pushchino}. 
These databases contain information about the FRB time, duration, energy spectrum, intensity, and redshift value (available only for some FRBs) .

Within the period of interest (December 2007 to June 2021), we selected $42$~FRBs with radio fluence values $\Phi_\mathrm{FRBi}$ above $40$~$\rm{(Jy~ms)}$.

The temporal correlation method is based on the fact that the radio fluence (which is supposed to be related to the neutrino fluence) from the time window of the most intense FRBs must be greater than the radio fluence from the window of background determination.
The expected total radio fluence from the given time window is proportional to its time span $\Delta t$ and the average radio flux from all FRBs $F_{all}$. In the windows of the most intense FRBs, the radio fluence will be increased by the average fluence $\Phi_{40}$.

In order to calculate the $F_{all}$ and $\Phi_{40}$ values, we used a power law index $\alpha=-1.4$ for cumulative fluence distribution and the all-sky FRB rate of $N_5=\rm{818~ sky^{-1} day^{-1}}$ above the fluence of 5 $\rm{(Jy~ ms)}$ obtained in \cite{Ami2021}.
A small additional contribution to $F_{all}$ from FRBs with the fluence $\leq 5 \rm{(Jy~ ms)}$ was made by adding $\simeq1200$ uniformly distributed fluences in the $(0 - 5) \rm{(Jy~ ms)}$ range.
As a result, the average flux from all FRBs turned out to be $F_{all}$ = $\rm{0.16~(Jy~ ms)s^{-1}}$ (or $7.0$~$\rm{(Jy~ ms)}$ per FRB) while the average fluence of FRBs with the fluence $\geq 40~ \rm{(Jy~ ms)}$ amounted to $\Phi_{40}$=74 $\rm{(Jy~ ms)}$.

Similar results were obtained from the analysis of specific FRBs.
According to the CHIME data~\cite{Ami2021}, the average radio fluence of all $536$~FRBs is $\Phi_{all} = 7.0$~$\rm{(Jy~ ms)}$ per FRB, while for $12$~FRBs with the fluence above $40$~$\rm{(Jy~ ms)}$, the average equals to $\Phi_{40} = 61.3$~$\rm{(Jy ~ ms)}$. 
Consequently, the average flux should be expressed as $F_{all} = \Phi_{all} N_{all} / T$~s$^{-1}$, where $N_{all}$ is the number of all-sky FRB events per day ($T = 24\rm{h} = 86400~\rm{s}$).

Thus, the excess of the expected neutrino events corresponding to the most intense FRB time intervals is defined by the factor $r = \Phi_{40} / (\Delta t F_{all})$.
Since in our study we chose the length of the time window $\Delta t = 2000$~s, this ratio turns out to be $r = 0.2$ of the average neutrino flux.


Independently, we analysed the Borexino data in coincidence with FRB~200428 from magnetar SGR~1935+2154 .
This event occurred on April 28, 2020 at the intragalactic distance of $9.5$~kpc, thus yielding a very high fluence of $1.5\times10^6$ $\rm{(Jy~ ms)}$.

The biggest redshift $z = 0.66$ was observed for FRB~190523.  
We have considered the coincidence time window $\Delta t = 2000$~s centered at the FRB observation time with a width of $\pm1000~$s covering a possible delay of sub-MeV neutrinos propagating at the sublight speed.
For a distance corresponding to the $z = 0.66$ redshift, the delay should reach $1000$~s in case of $0.6$~MeV neutrinos with a rest mass of $70$~meV, which is the upper limit on the heaviest neutrino mass state from the Planck 2015 data and oscillation mass squared differences~\cite{Ago2017b}. 


The FRB arrival time could have its own delay associated with propagation through intergalactic plasma.
The delay depends on the registered dispersion measure and the frequency $\omega$ at which the signal was recorded as $\omega^{-2}$~\cite{Lor2007,Pop2018,Lor2004}.

Only the Pushchino telescope~\cite{pushchino} operating at a sufficiently low frequency of  $109-113$~MHz was able to register FRB~160920 with a significantly large delay of $620$~s. 
Among $11$~FRBs detected by the Pushchino telescope, only $6$~FRBs have a signal delay exceeding $100$~s. 
Higher working frequencies (up to $1$~GHz) of all other telescopes result in delays below $100$~s. 
However, these delays were calculated and taken into account in our analysis.

All selected FRBs had the data taking time above $95$\% of the corresponding time interval $\Delta t$.
\begin{figure}
	\includegraphics[width=8cm, height=8cm]{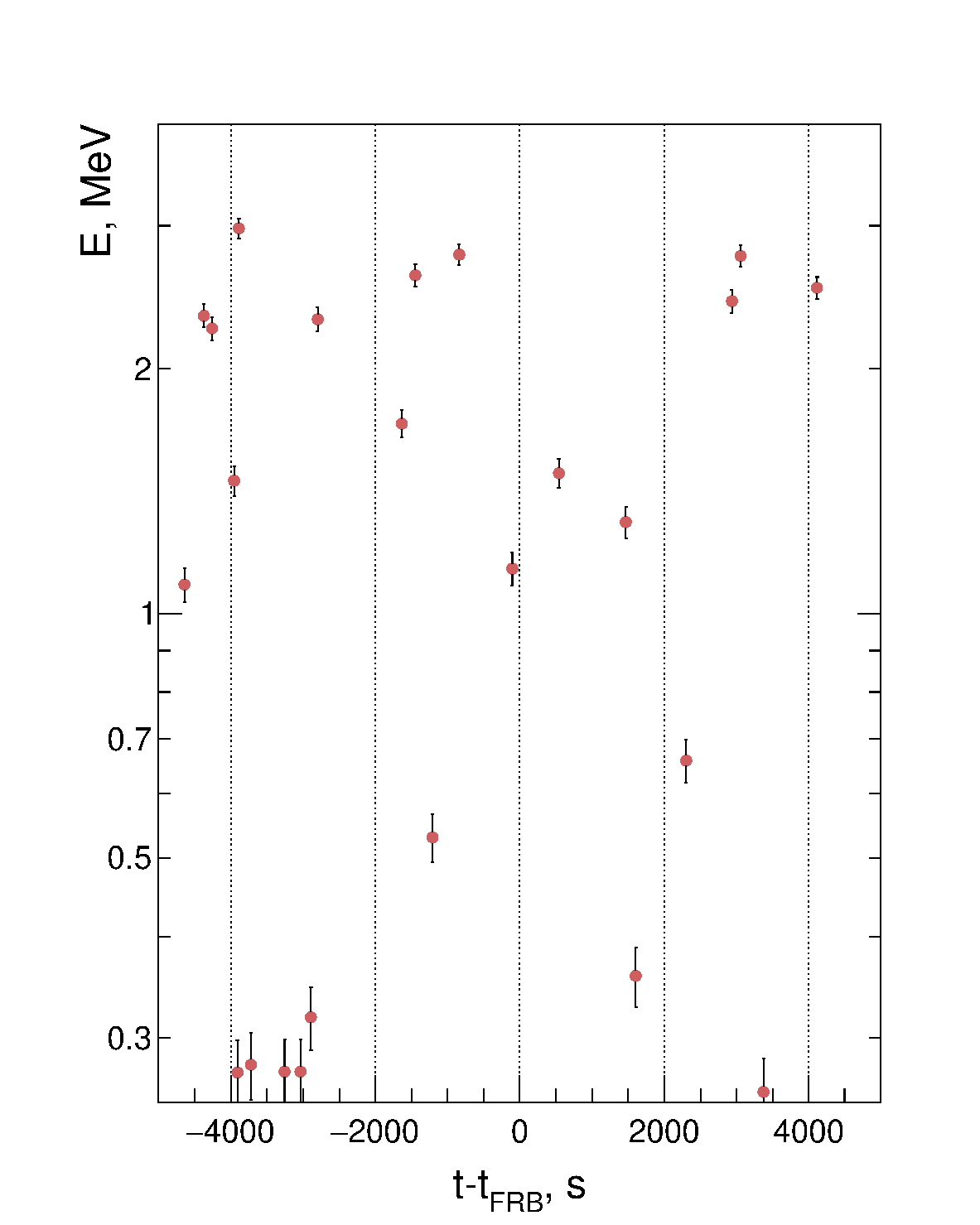}
	\caption{Borexino events with an energy above $0.25$~MeV occurring within $\pm5000$~s of FRB~200428 detection time.}\label{fig:1} 
\end{figure}
\begin{figure}
	\includegraphics[width=9cm, height=10cm]{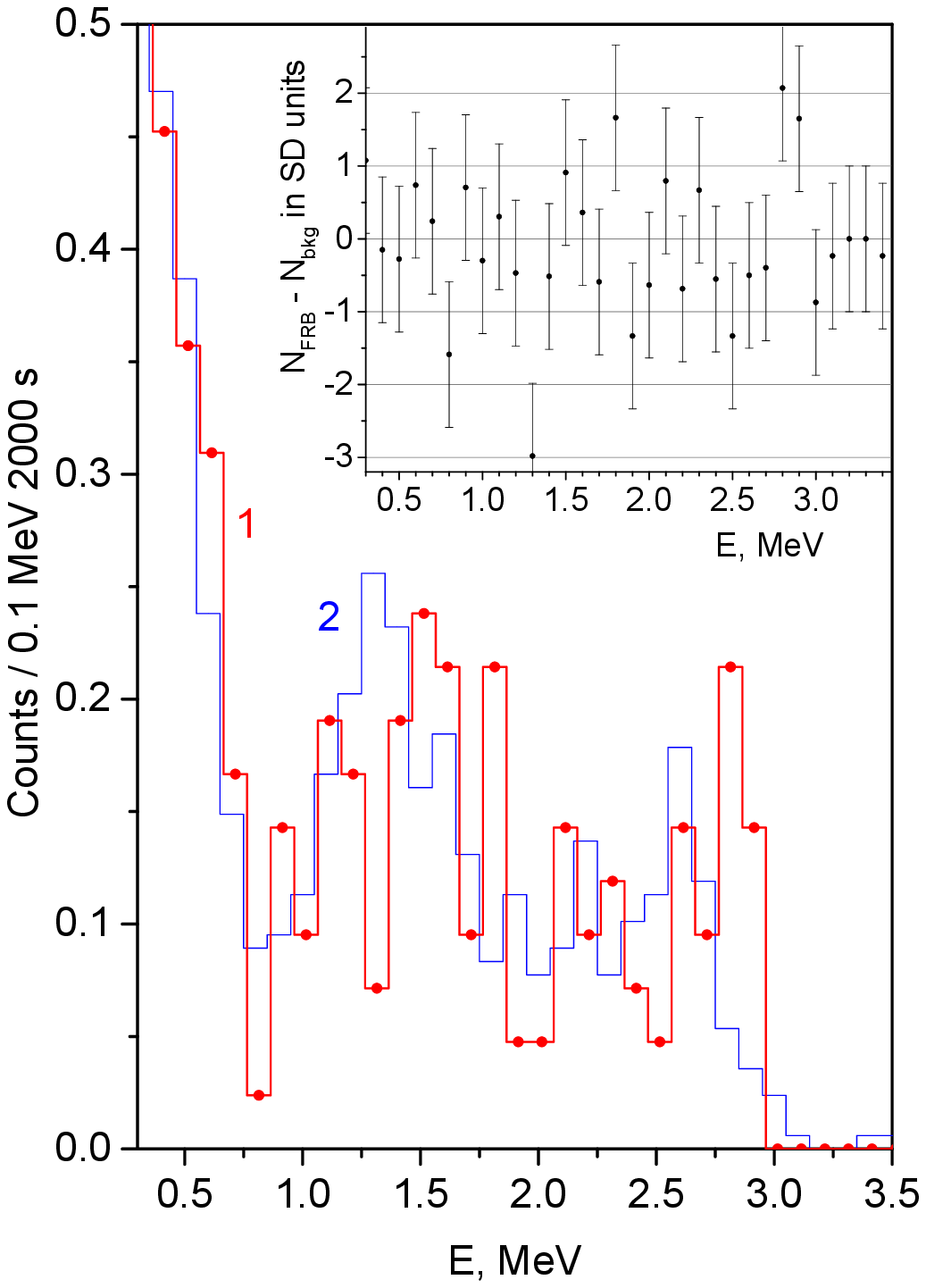}
	\caption{Borexino energy spectrum of singles in correlation with FRBs in the $\pm1000$~s time window (Line 1 with dots). Line 2 shows the normalized background spectrum measured in $\left[{-5000} \dots {-1000}\right]$~s and $\left[{1000} \ldots {5000}\right]$~s intervals. The inset shows the difference between these spectra  in terms of standard deviations (SD).} \label{fig:2}
\end{figure}
\begin{figure}
	\includegraphics[width=8cm, height=8cm]{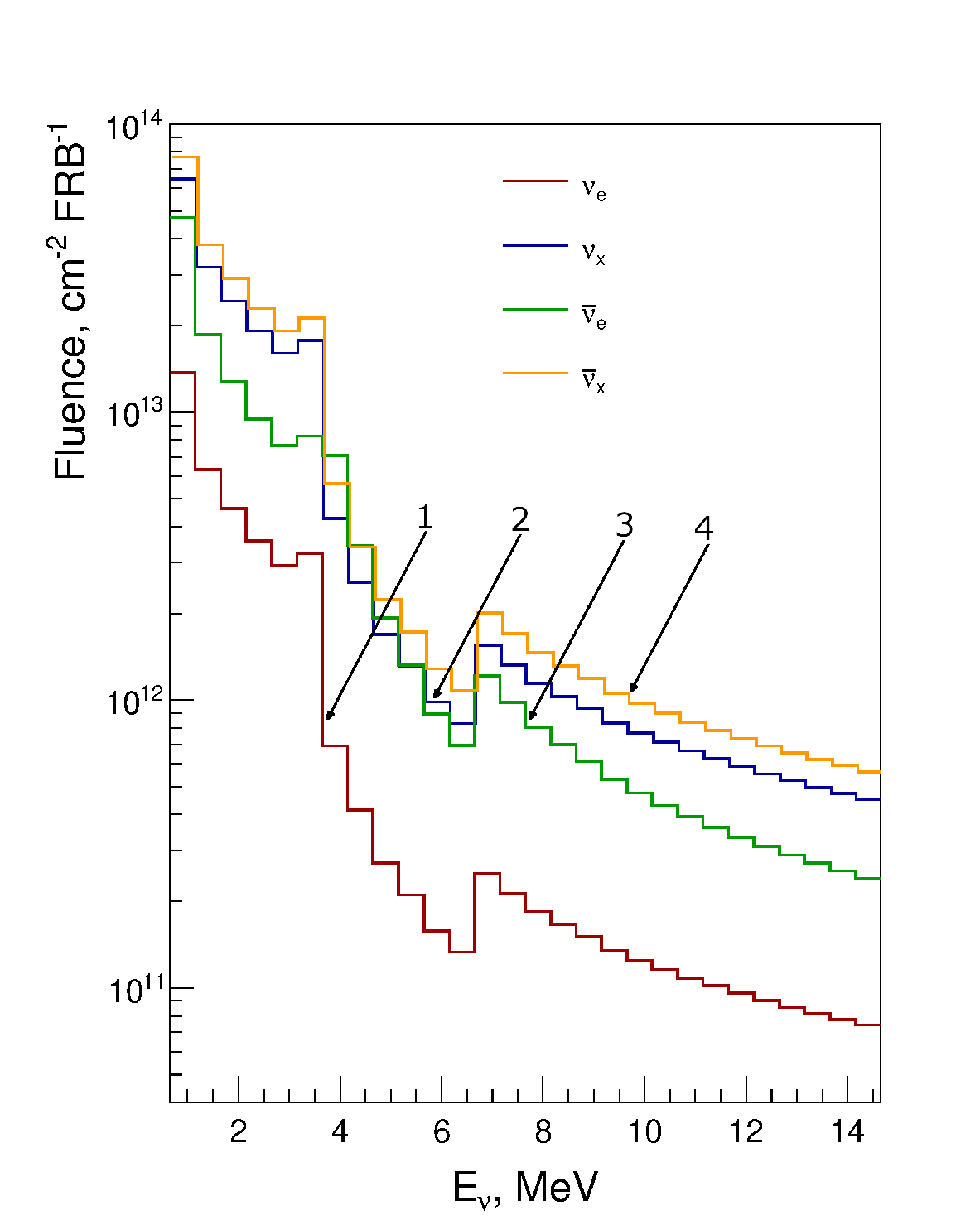}
	\caption{90\% C.L. upper limits on mono-energetic neutrino fluences obtained through the temporal correlation analysis for 42 most intensive FRBs with the fluence $\Phi_{\rm{FRB}} \geq 40~\rm{Jy~ms}$: 1~--~$\nu_e$, 2~--~$\nu_{\mu,\tau}$, 3~--~$\bar{\nu}_e$, 4~--~$\bar{\nu}_{\mu,\tau}$.}\label{fig:3}
\end{figure}

\subsection{Neutrino Spectra}

As noted above, the origin of FRBs is largely unknown.
It is not even clear whether all of the observed radio bursts belong to the same type of physical processes. 
Among many different FRB models, there are several, such as neutron star mergers or supramassive neutron star collapses, which predict collateral neutrino and axion radiation~\cite{Pop2018,Pla2020}. 
In case of high-energy $\mathrm{GeV} - \mathrm{PeV}$ neutrinos produced in hadronic processes, one would expect the power law spectrum $\sim E_\nu^{-2}$. There is no reliable theory of the low-energy part of the FRB neutrino emission spectrum. 

We calculated fluence limits on two different kinds of possible neutrino spectra: the monoenergetic line and the SN low-energy continuous spectrum.
The latter was assumed as a quasi-thermal spectrum with the mean energy $\langle E \rangle$ and deviation from thermal distribution characterized by the pinching parameter $\alpha = 3$ for all neutrino flavors ($\nu_x, \bar{\nu}_x$)~\cite{Tam2012,Luj2014,Mir2016}.
The emitted neutrino spectrum $S(E_\nu)$ depends on the neutrino energy $E_\nu$ as:
\begin{equation}\label{eq:5}
    S(E_\nu) \sim (E_\nu/T)^\alpha e^{(-E_\nu/T)},
\end{equation} 
where $T = \langle E \rangle / (\alpha + 1)$ is the effective temperature, which was considered to be the same for all neutrino flavors.

\subsection{Temporal Correlations for the Most Intensive FRBs}

The goal of this analysis was to search for an excess of the selected events above the measured background, in coincidence with FRBs in a time window of $\Delta t = 2000~$s centered at the FRB arrival time.
We calculated the overall number of candidate events above $250$~keV in the $\Delta t$~interval, which met the requirements for selection cuts of the described data.

For reference, the Borexino events with an energy above 0.25 MeV produced within $\pm 5000~$s of the most intensive Galactic FRB 200428 detection time are shown in Figure~\ref{fig:1}.
The closest events, with energies of $1.13$~MeV and $1.49$~MeV, occurred $105$~s before and $539$~s after the FRB arrival, respectively. 
There were only three events in the $\pm1000$~s interval with the energy in the $0.25 - 15$~MeV range, while $4.4 \pm 0.1$ solar neutrino and background events were obtained within the same time/energy window from the weekly run containing  FRB~200428.
All detected events were in agreement with the expected solar neutrino and background count rate.

Figure~\ref{fig:2} shows the energy spectrum measured for the integrated time exposure $N_\mathrm{FRB} \times \Delta t$ in the $250~\mathrm{keV} - 3.5~\mathrm{MeV}$ energy range.
There is only a single $6.8$~MeV event outside this energy interval.
For comparison, the same Figure~\ref{fig:2} contains the background spectrum measured in two adjacent time intervals, $\left[ {-5000}\ldots {-1000}\right]$~s and $\left[{1000} \ldots {5000} \right]$~s.
No statistically significant excess of the difference between these spectra for any energy interval was observed.

We calculated the upper limits on fluences $\Phi_{\nu_x,\bar{\nu_x}}$ for monoenergetic (anti)neutrinos with the energy $E_\nu$ as:
\begin{equation}\label{eq:6}
    \Phi_{\nu_x,\bar{\nu_x}} =
        \frac {N_{90}(E_\nu, n_\mathrm{obs}, n_\mathrm{bkg})} {r N_e \sigma(E_\mathrm{th}, E_{e_\mathrm{max}})},
\end{equation}
where $N_{90} (E_\nu, n_\mathrm{obs}, n_\mathrm{bkg})$ is the 90\%~C.L. upper limit for the number of FRB-correlated events in the $(E_\mathrm{th}, E_{e_\mathrm{max}})$ interval per single FRB, $N_e$ is the number of electrons in $145$~t of the Borexino scintillator.
The factor $\sigma(E_\mathrm{th}, E_{e_\mathrm{max}})$ represents the cross section for detected neutrinos $(\nu_{x}, \bar{\nu_x})$ with the energy $E_{\nu}$ without oscillations while recoil electrons are detected in the interval ($E_\mathrm{th}, E_{e_\mathrm{max}}$)~\cite{Ago2017a}. 
The recoil electron detection efficiency was taken as 1, with the accuracy corresponding to the precision of the fiducial volume definition ($\simeq1\%$).

The numerator $N_{90}(E_\nu,n_\mathrm{obs},n_\mathrm{bkg})$ was calculated taking into account the statistical compatibility between the correlated and uncorrelated FRB spectra and assuming Poisson statistics for the number of detected events in the energy interval ($E_\mathrm{th}, E_{e_\mathrm{max}}$).
Here, $n_\mathrm{obs}$ and $n_\mathrm{bkg}$ denote overall numbers of events in the energy interval ($E_\mathrm{th}, E_{e_\mathrm{max}}$) detected in the time periods $N_\mathrm{FRB} \times \Delta t_\mathrm{obs}$ and $N_\mathrm{FRB} \times \Delta t_\mathrm{bkg}$ normalized by their respective times. 
The intervals $\Delta t_\mathrm{obs}$=2~ks and $\Delta t_\mathrm{bkg}$ =18~ks were taken as $\pm 1000$~s and sum of $\left[ -10000\ldots -1000\right]$~s and $\left[1000\ldots 10000\right]$~s, correspondingly. 
The longer interval for background detection was chosen in order to reduce the error of $n_\mathrm{bkg}$ that plays an important role in the Feldman-Cousins method.
The value $n_\mathrm{bkg}$ was normalized by the overall time ratio taking into account the actual live time of the detector within these time windows. 

The procedure was repeated for neutrino energies $E_{\nu}$ from 0.5~MeV to 15~MeV in increments of 0.5 MeV. In order to have the best ratio of the expected effect with respect to the background and taking into account the shape of the spectrum (Figure~\ref{fig:2}) in addition to the 0.25 MeV threshold, the 3.0 MeV threshold was used for higher neutrino energies. The upper limits on neutrino and antineutrino fluences of different flavors normalized per single FRB are shown in Figure ~\ref{fig:3}. The jump in the upper limit at energies above 7 MeV is associated with the inclusion of the above-mentioned 6.8~MeV event in the analysis.

These are the first constraints on the MeV neutrino fluxes obtained from the neutrino-electron scattering reaction. 
The average radio fluence of the most intensive FRBs and the limits on the neutrino fluence shown in Figure~\ref{fig:3} allow us to obtain the limit on the ratio $(\Phi_\nu / \Phi_\mathrm{FRB})$ in $(\nu~ \rm{cm^{-2} / Jy~ ms})$ units.

Since there is no reliable model for the low-energy neutrino spectrum for FRBs, we perform calculations of neutrino emission from a supernova collapse~\cite{Tam2012,Luj2014,Mir2016}.
Assuming quasi-thermal distributions~(\ref{eq:5}) with a mean energy $\langle E \rangle = 15.6$~MeV and the parameter $\alpha = 3$ and integrating over the analyzed electron recoil energy interval $0.25 - 15.0$~MeV, we get the limits on the total electron neutrino fluence per single FRB: $\Phi(\nu_e) \leq 3.69 \times 10^{10} \rm{cm^{-2}}$ (90\%~C.L.) that is about three times weaker than the limit obtained for monoenergetic neutrinos with the same energy.
The values of the limits on other neutrino flavors obtained from the $(\nu,e)$-scattering channel, as well as from the IBD reaction for $\bar{\nu}_e$, are given in Table~\ref{tab:1}. 

We also calculated the upper limit on the electron antineutrinos ($\bar{\nu}_e$) fluence using the IBD reaction from the relation (\ref{eq:6}) but replacing $N_e$ with the number of protons $N_p$, considering cross-section of IBD reaction and using the data acquired between December 2007 and October 2017 \cite{Ago2021}. 
The data period considered includes only 19 FRB events above the threshold $40$~$\rm{(Jy~ ms)}$.
No IBD events were observed in the $\pm 1000$~s interval around the selected FRBs and the expected background was almost zero~\cite{Ago2021,Ago2020b} that allowed us to use the conservative value of $N_{90}(E_\nu,n_\mathrm{obs},n_\mathrm{bkg})$ = 2.44 in the analysis~\cite{Fel1998}. Since the cross section of IBD reaction is about two orders of magnitude larger than $(\nu,e)$-scattering cross-sections at given neutrino energies, and the background level is smaller, the most stringent upper limits have been obtained for the fluence of electron antineutrinos.

\begin{table}[ht]
\caption{Upper limits on fluences per single FRB for all neutrino flavors obtained from the temporal correlation analysis in $10^{9}~\rm{cm^{-2}}$ units (90\%~C.L.) calculated for monoenergetic neutrinos and the SN spectrum with $\langle E \rangle = 15.6$~MeV.}
\label{tab:1}
\begin{tabular}{|c|c|c|c|c|c|}
\hline
$E_{\nu}$ & ${\Phi}_{\nu_e}$ & ${\Phi}_{\bar{\nu}_{e}}$ & ${\Phi}_{\nu_{\mu,\tau}}$ & ${\Phi}_{\bar{\nu}_{\mu,\tau}}$ & IBD\\
\hline
2 & $4620$ & $12750$  & $ 24250$ & $29000$ & $2475$ \\
6   & $157$ & $890$ & $890$ & $1280$ & $40.7$ \\ 
10   & $125$ & $475$ & $770$ & $970$ & $12.2$ \\ 
14   & $77.5$ & $255$ & $474$ & $590$ & $5.88$ \\ 
$<15.6>$ & $157$ & $367$ & $900$ & $1070$ & $11.6$ \\ 
\hline
\end{tabular}
\end{table}

\begin{figure}
	\includegraphics[width=9cm]{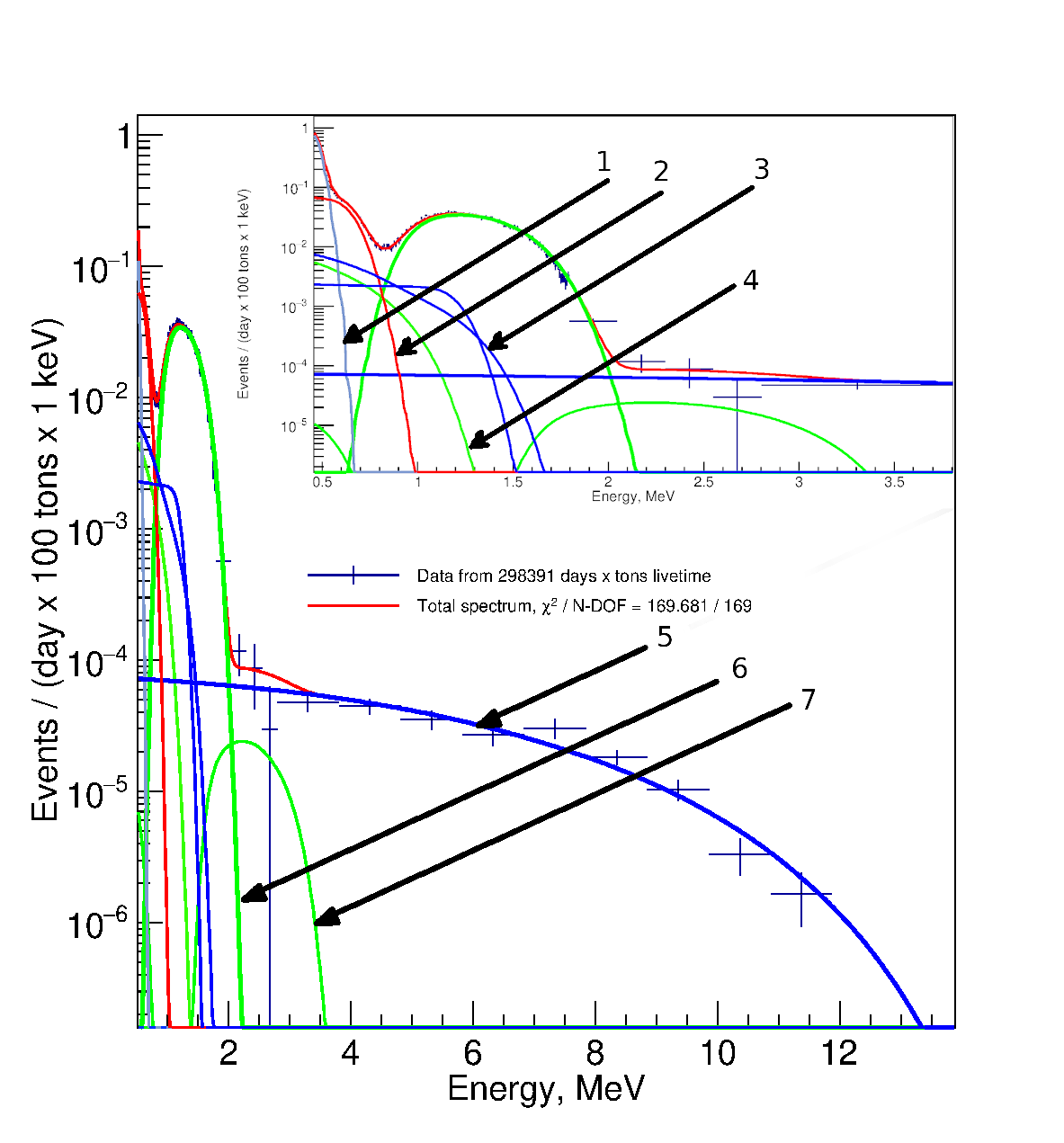}
	\caption{Spectral fit of the selected Borexino data after statistical subtraction of external backgrounds. The spectral components considered are: 1~--~$^{210}$Po $\alpha$-peak, 2~--~recoiled electrons from the solar $^7$Be, 3--~recoiled electrons from the solar CNO and $pep$-neutrinos, 4~--~$^{210}$Bi $\beta$-spectrum, 5~--~solar $^8$B neutrino electron recoils, 6~--~$^{11}$C $\beta^+$-decay, 7~--~$^{10}$C $\beta^+$-decay. The inset shows the data in the energy range [0.5~--~4]~MeV. The fit shows a good statistical agreement with this spectral composition model and with the fluxes that are in agreement with~\cite{Bel2014,Ago2018}.  }
	\label{fig:4}      
\end{figure}

\subsection{Limits on the $\nu_{e,\mu,\tau}$ and $\bar{{\nu}}_{e,\mu,\tau}$ Fluences from the Spectral Fit}

Figure~\ref{fig:4} shows the Borexino spectrum in the range $0.5 - 14$ MeV after applying data selection cuts from Section 4.2 for $298.39$~$\mathrm{kt ~ day}$ statistics or $2058$~days of live time.
The spectrum is dominated by $^{14}$C in the region below $250$~keV (outside the drawing range), electron recoil from solar $^7$Be neutrinos in the $0.25 - 1$~MeV interval, by cosmogenic $^{11}$C in the $1 - 2$~MeV region, and by $^8$B solar neutrinos above $2$~MeV. 
The background used in the fit procedure was described with the actual spectral components, such as $^{210}$Po, $^{85}$Kr, solar neutrino recoil electrons, $^{11}$C, and $^{10}$C. 
External gammas were statistically subtracted and thus were not included in the fit that was performed with the standard $\chi^2$ likelihood function.
The detector energy response was described according to~\cite{Smi2008} with additional empiric consideration of light quenching and Cherenkov radiation emission. 
The energy scale calibration and light quenching for $\alpha$- and $\beta^\pm$-particles were left as free parameters of the likelihood function~\cite{Bel2014}.

The obtained count rates of all spectral components are in a good statistical agreement with previous publications~\cite{Bel2014,Ago2018,Ago2020a,Ago2020e}, but have larger uncertainties caused by the statistical subtraction procedure.

The additional component responsible for the potential FRB signal was added to the background and had a spectral shape of the monoenergetic line with the energy $E_\nu$ or the supernova spectrum given by~(\ref{eq:5}) with different $\langle E \rangle$ values.
The limit on the number of events from the additional component was derived using the $\chi^2$ profile as the value that corresponds to increase by $(1.64)^2$ with respect to the minimal value or the value at zero count rate in cases when the best-fit count rate value turned out to be negative.
This limit $N_{90}(E_\nu)$ corresponds to a confidence level of 90\% and could be converted into limit on the FRB fluence as:
\begin{equation}\label{eq:7}
   \Phi = \frac {N_{90}(E_\nu)} {N_e \sigma(E_\mathrm{th},E_\nu)},
\end{equation}
where $N_e$ is the number of electrons in the FV scintillator and $\sigma(E_\mathrm{th},E_\nu)$ is the $(\nu, e)$-scattering cross section. 

\begin{figure}
	\includegraphics[width=9cm, height=10cm]{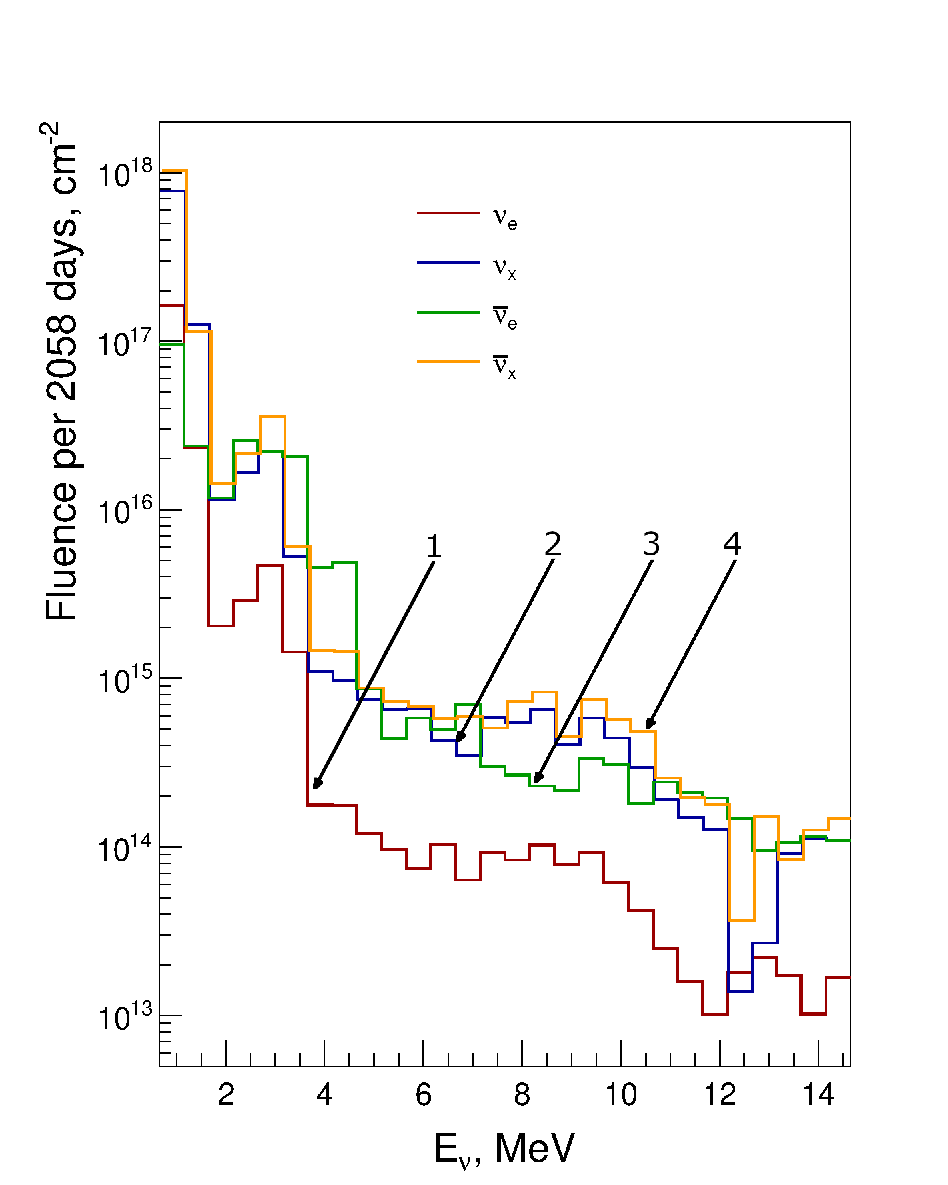}
	\caption{Upper limits on the fluences of monoenergetic  $\nu_{e,\mu,\tau}$ and $\bar{\nu}_{e,\mu,\tau}$ obtained from the spectral fit (90\%~C.L.): 1~--~$\nu_e$, 2~--~$\nu_{\mu,\tau}$, 3~--~$\bar{\nu}_e$, 4~--~$\bar{\nu}_{\mu,\tau}$.}\label{fig:5}
\end{figure}

The limits on the total fluence (time-integrated neutrino flux) during $2058$~days of measurements obtained with this procedure for monoenergetic neutrinos of all flavors are shown in Figure~\ref{fig:5}. 
Assuming an expected all-sky FRBs rate of $N_{all} = 2\times10^3$ per day~\cite{Pet2019,Bha2018}, the limits per single FRB are $4\times10^6$ times stronger than those shown in Figure~\ref{fig:5}.
Comparing the limits obtained from the temporal correlation analysis (Figure~\ref{fig:3}) and from the spectral fit (Figure~\ref{fig:5}), one can see that the latter are about $(1-3)\times10^4$ times more stringent although they depend on the assumed all-sky FRB count rate $N_{all}$.

Assuming that the neutrino fluence is proportional to the radio one and taking into account that, as shown in Section~4.1, the average fluence from all flares (7.0 (Jy~ ms) per FRB) is an order of magnitude less than the average fluence $\Phi_{40}$ of the most intense ones, the limits on neutrino per radio fluence obtained from spectral analysis are only $(1-3)\times10^3$ times stronger than those obtained from the temporal analysis.

\begin{figure}
	\includegraphics[width=9cm, height=10cm]{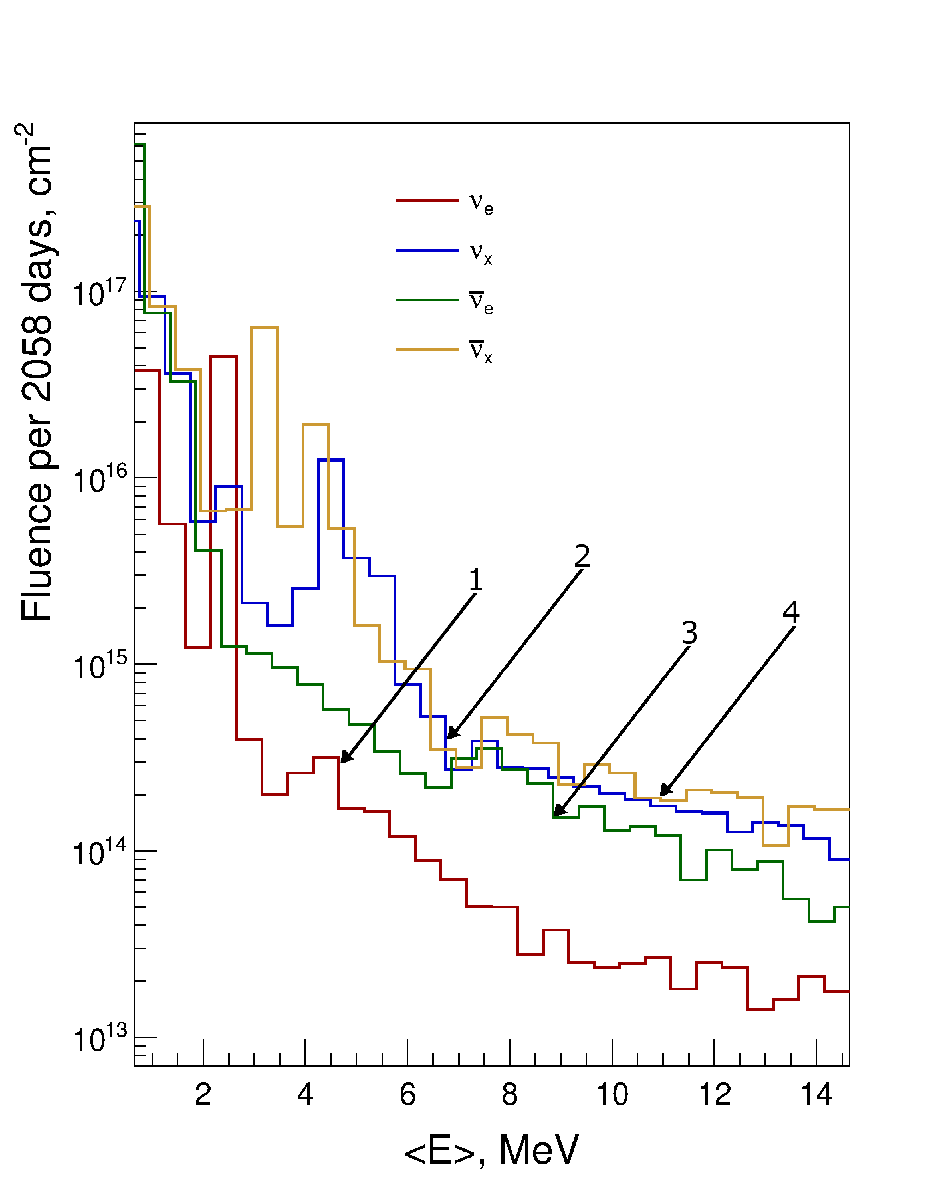}
	\caption{Upper limits on the fluences of $\nu_{e,\mu,\tau}$ and $\bar{\nu}_{e,\mu,\tau}$ with supernova neutrino spectra obtained from the spectral fit (90\%~C.L.): 1~--~$\nu_e$, 2~--~$\nu_{\mu,\tau}$, 3~--~$\bar{\nu}_e$, 4~--~$\bar{\nu}_{\mu,\tau}$.}\label{fig:6}
\end{figure}

The same analysis was applied to the neutrino spectrum from a supernova given by eq.~(\ref{eq:5}).
Figure~\ref{fig:6} shows the 90\%~C.L. upper limits on fluences for supernova neutrino spectra with different mean energies $\langle E \rangle$. 
The correlation of the expected neutrino signal with the components of the Borexino spectrum (Figure~\ref{fig:4}) leads to the significant variations in the fluence upper limits observed in Figures~\ref{fig:5},~\ref{fig:6} depending on the neutrino energy $E_{\nu}$ and $\langle E \rangle$.
For a value of $\langle E \rangle = 15.6$~MeV, the upper limit per total fluence of $\nu_e$ is $\Phi_{\nu_e} \leq 1.66\times10^{13}$~$\mathrm{cm}^{-2}$. 
Taking into account the number of FRBs expected in $2058$~days, the obtained limit turns out to be $9\times10^3$ times more stringent than the one obtained from the temporal analysis (Table~\ref{tab:1}).

\subsection{Limits Obtained from the Absence of the $(\nu,e)$-Events above 13.6 MeV}

As for the neutrinos with higher energies, the limits on $\nu_{e,\mu,\tau}$ and $\bar{\nu}_{e,\mu,\tau}$ were alternatively obtained using the fact that no events were observed above $13.6$~MeV. 
The right boundary of the interval for the analysis was set to $16.8$~MeV in accordance with the verified energy calibration of the data acquisition system.
The different values of the monoenergetic neutrino energy $E_\nu$ and the supernova neutrino mean energy $\langle E \rangle$, the expected spectra of recoil electrons and the number of events $N_x$ in the interval $13.6 - 16.8$~MeV were used in the calculations. 

\begin{figure}
	\includegraphics[width=9cm, height=10cm]{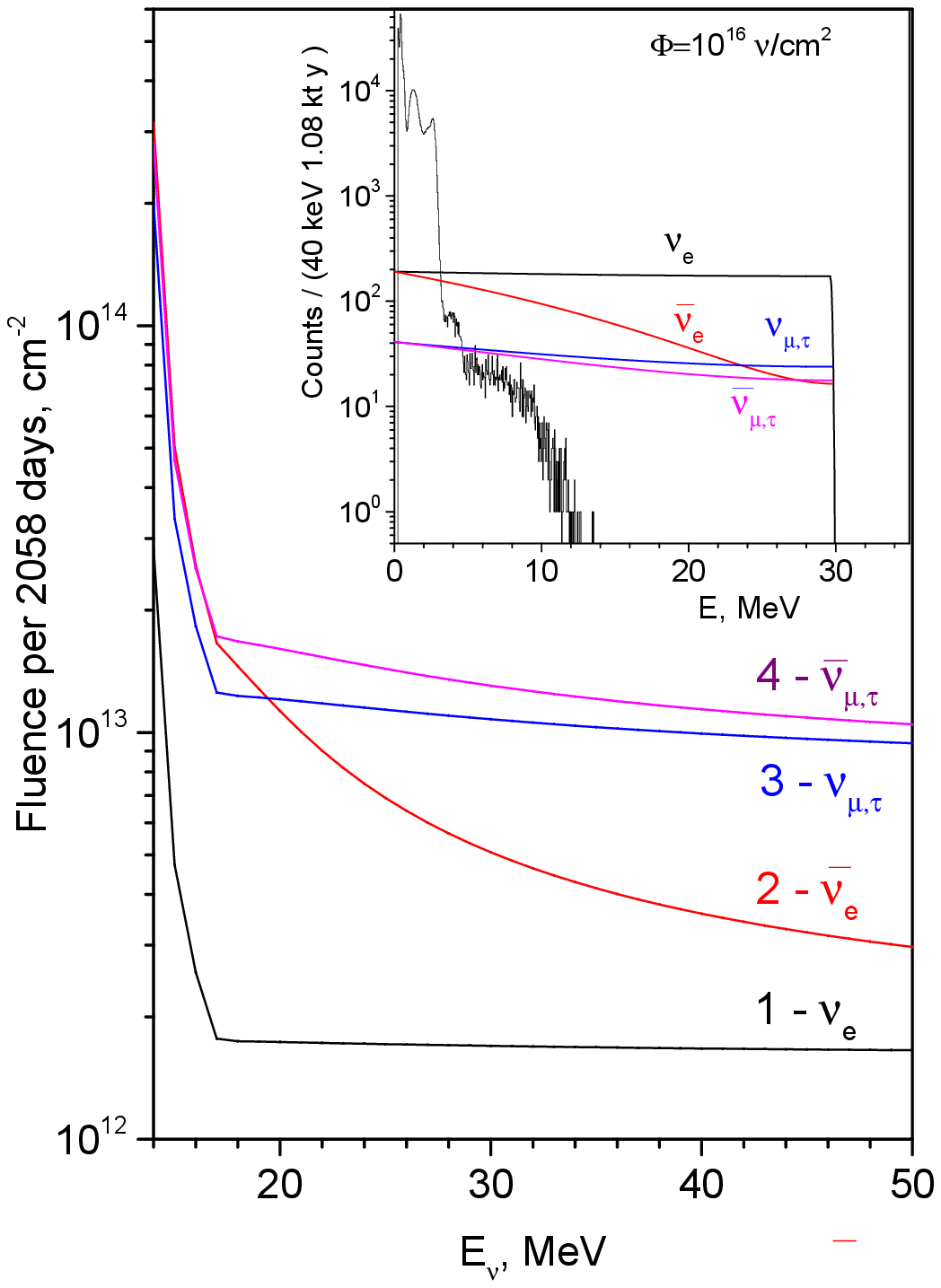}
	\caption{The upper limits on the fluences of monoenergetic $\nu_{e,\mu,\tau}$ and $\bar{{\nu}}_{e,\mu,\tau}$ with 1 MeV step (90\%~C.L.). The inset shows the Borexino data and the expected $(\nu_x, e)$-scattering spectra for $30$~MeV neutrinos.}\label{fig:7}
\end{figure}

\begin{figure}
	\includegraphics[width=9cm, height=10cm]{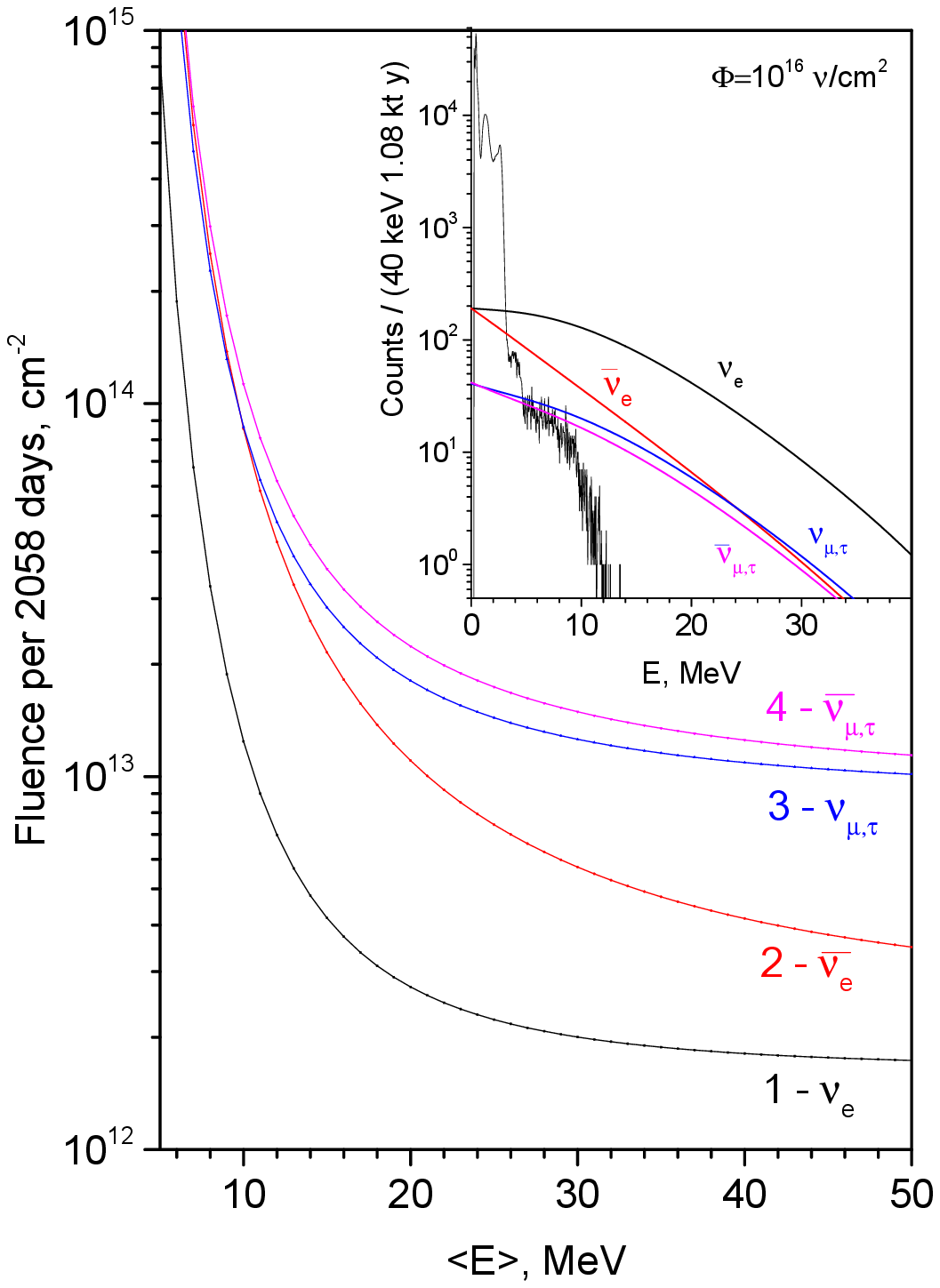}
	\caption{The same as in Figure~\ref{fig:7}, but for the supernova spectra. The inset shows the expected electron recoil spectra for the case $\langle E \rangle = 15.6$~MeV.}\label{fig:8}
\end{figure}

According to the Feldman-Cousins approach to the case of no observed events with the conservative zero background, the $90$\%~C.L. upper limit  in the $13.6 -16.8$~MeV interval is $N_\mathrm{lim} = 2.44$.
The relation $N_x \leq N_\mathrm{lim}$ was converted into the obtained fluence limits for all neutrino flavors given in Figure~\ref{fig:7} for monoenergetic neutrinos and in Figure~\ref{fig:8} for supernova neutrinos. 
The figures \ref{fig:7} and \ref{fig:8} also show the expected spectra of recoil electrons for the $(\nu,e)$-elastic scattering reactions for $E_{\nu_x} = 30$~MeV and $\langle E \rangle = 15.6$~MeV. 

The region of the neutrino energy up to $50$~MeV was chosen in accordance with the characteristic energies of neutrinos appearing in the pion decay at rest.
A sharp decrease in sensitivity to the neutrino fluence at the energies below $17$~MeV occurs when the neutrino energy enters the $13.6 - 16.8$~MeV interval. 
At higher energies, the limit becomes almost constant since the $(\nu,e)$-scattering cross section is proportional to $E_\nu$ and the spectrum of recoil electrons weakly depends on the electron energy (Figure~\ref{fig:7}).

\begin{table}[ht]
\caption{Borexino $90$\%~C.L. upper limits for the total FRB fluences of all neutrino flavours, obtained through the study of $(\nu,e)$ elastic scattering of monoenergetic neutrinos  ($2058$~d) and IBD reaction ($2485$~d). $E_{\nu}$ is given in MeV units, ${\Phi}_{\nu_x,\bar{\nu}_x}$ -- in $10^{12}~\rm{cm^{-2}}$ units. }
\label{tab:2}

\begin{tabular}{|c|c|c|c|c|c|}
\hline
$E_{\nu}$ & ${\Phi}_{\nu_e}$ & ${\Phi}_{\bar{\nu}_{e}}$ & ${\Phi}_{\nu_{\mu,\tau}}$ & ${\Phi}_{\bar{\nu}_{\mu,\tau}}$ & IBD\\
\hline
2 & $2040$ & $11700$  & $11400$ & $14200$ & $30.1 $ \\
6   & $74.8$ & $584$ & $662$ & $678$ & $0.173 $ \\
10   & $61.6$ & $310$ & $444$ & $574$ & $0.027 $ \\
14   & $10.2$ & $116$ & $112$ & $126$ & $0.013 $ \\
18   & $ 1.74 $ & $ 14.5 $ & $ 12.3 $ & $ 16.8 $ & $0.008 $ \\
30   & $ 1.70 $ & $ 5.08 $ & $ 10.8 $ & $ 13.0 $ & $ - $ \\
50   & $ 1.66 $ & $ 2.97 $ & $ 9.41 $ & $ 10.5 $ & $ - $ \\
\hline
\end{tabular}
\end{table}

The results of the spectral fit and the case of no observed events are also shown in Table~\ref{tab:2} for monoenergetic neutrinos. 
One can see the validity of using different analysis methods.
The spectral fit gives a better sensitivity to lower neutrino energies in the $0.5 -14.0$~MeV range while the second method allows expanding the neutrino energy range up to $50$~MeV.

In case of the supernova neutrino spectrum, the fluence constraints based on the absence of events above $13.6$~MeV becomes stronger than the monoenergetic neutrino limit starting from the mean energies above $\langle E \rangle \geq 10$~MeV. 
The resulting most stringent limits are given in Table~\ref{tab:3}.

\begin{table}[ht]
\caption{Borexino 90\%~C.L. upper limits for the total FRB fluences obtained through the study of $(\nu,e)$ scattering ($2058$~d) and the IBD reaction ($2485$~d) with supernova neutrino spectra. The mean energy  $\langle E \rangle$ is given in MeV units, ${\Phi}_{\nu_x,\bar{\nu}_x}$ -- in $10^{12}~\rm{cm^{-2}}$ units. }
\label{tab:3}

\begin{tabular}{|c|c|c|c|c|c|}
\hline
$\langle E \rangle$ & ${\Phi}_{\nu_e}$ & ${\Phi}_{\bar{\nu}_{e}}$ & ${\Phi}_{\nu_{\mu,\tau}}$ & ${\Phi}_{\bar{\nu}_{\mu,\tau}}$ & IBD\\
\hline
2 & $1220$ & $4060$  & $5840$ & $6640$ & $528$ \\
6   & $119$ & $259$ & $779$ & $943$ & $0.117 $\\
10   & $12.4$ & $85.6$ & $86.6$ & $113$ & $0.036 $\\
14   & $4.8$ & $26.1$ & $32.8$ & $41.8$ & $0.030 $\\
18   & $3.11$ & $13.8$ & $20.8$ & $26.0$ & $0.034$\\
30   & $2.00$ & $5.72$ & $12.6$ & $14.9$ & $0.073$\\
50   & $1.73$ & $3.49$ & $10.1$ & $11.4$ & $0.245$\\
\hline
\end{tabular}
\end{table}

\subsection {Limits on the $\bar{\nu}_e$-Fluence from the IBD Reaction}
As already mentioned, electron antineutrinos can be also detected in Borexino via the neutron inverse $\beta$-decay (IBD) reaction on protons with an energy threshold of $1.8$~MeV.
The cross section of this process is much higher than the one for $(\bar{\nu}_e,e)$ elastic scattering. Additionally, the IBD offers a unique signature given by temporal and spatial coincidence of two correlated events associated with the detection of the positron and the neutron. 
The prompt positron event with a visible energy of $E_\nu - 0.784$~MeV accompanied by $\gamma$-rays from neutron capture mostly on protons or carbon nuclei with a small probability.
As a result, the rate of the events selected as IBD candidates is much lower with respect to the rate of single electron-like events.

The present study of the electron antineutrino flux associated with FRBs is based on the data acquired between December 2007 and October 2017.
The procedure of IBD events selection and the energy spectrum of prompt positron events are described in detail in~\cite{Ago2021}.
After the application of all selection cuts the total live time decreases to $2485$~days and the final efficiency of IBD reaction detection turns out to be $0.85$ ~\cite{Ago2021}. 
All electron antineutrino candidates were identified with the main DAQ~system as well as the FADC system that provides a linear dynamic range up to $\sim 50$~MeV.
In this analysis, we used the same $16.8$~MeV upper boundary of the antineutrino energy range as in the case of the $(\nu,e)$-scattering analysis.

\begin{figure}
	\includegraphics[width=9cm, height=10cm]{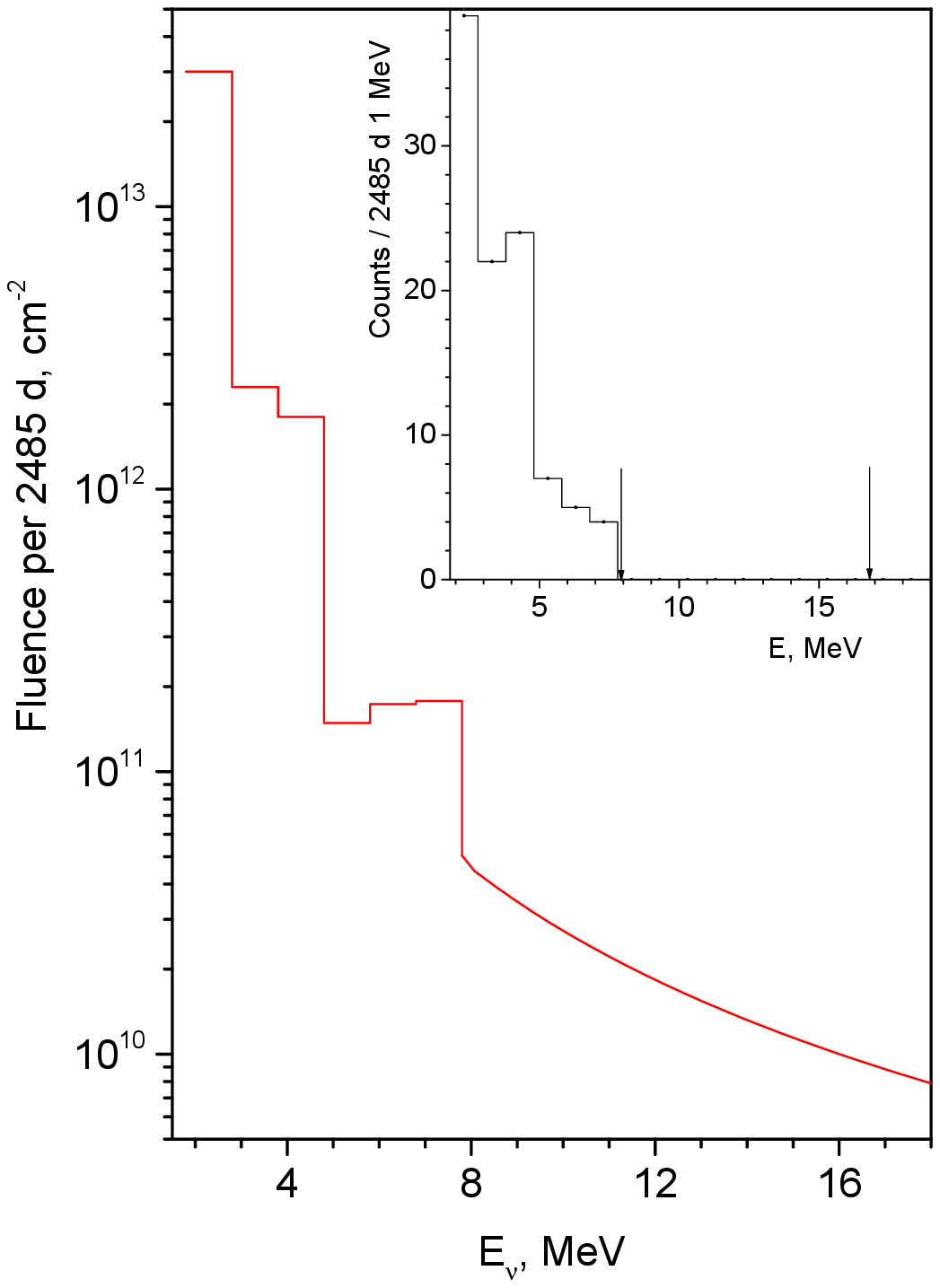}
	\caption{Upper limits on the fluences of monoenergetic $\bar{\nu}_e$ (90\%~C.L.). The inset shows the Borexino spectrum of $\bar{\nu}_e$-like events~\cite{Ago2021}.}\label{fig:9}
\end{figure}

\begin{figure}
	\includegraphics[width=9cm, height=10cm]{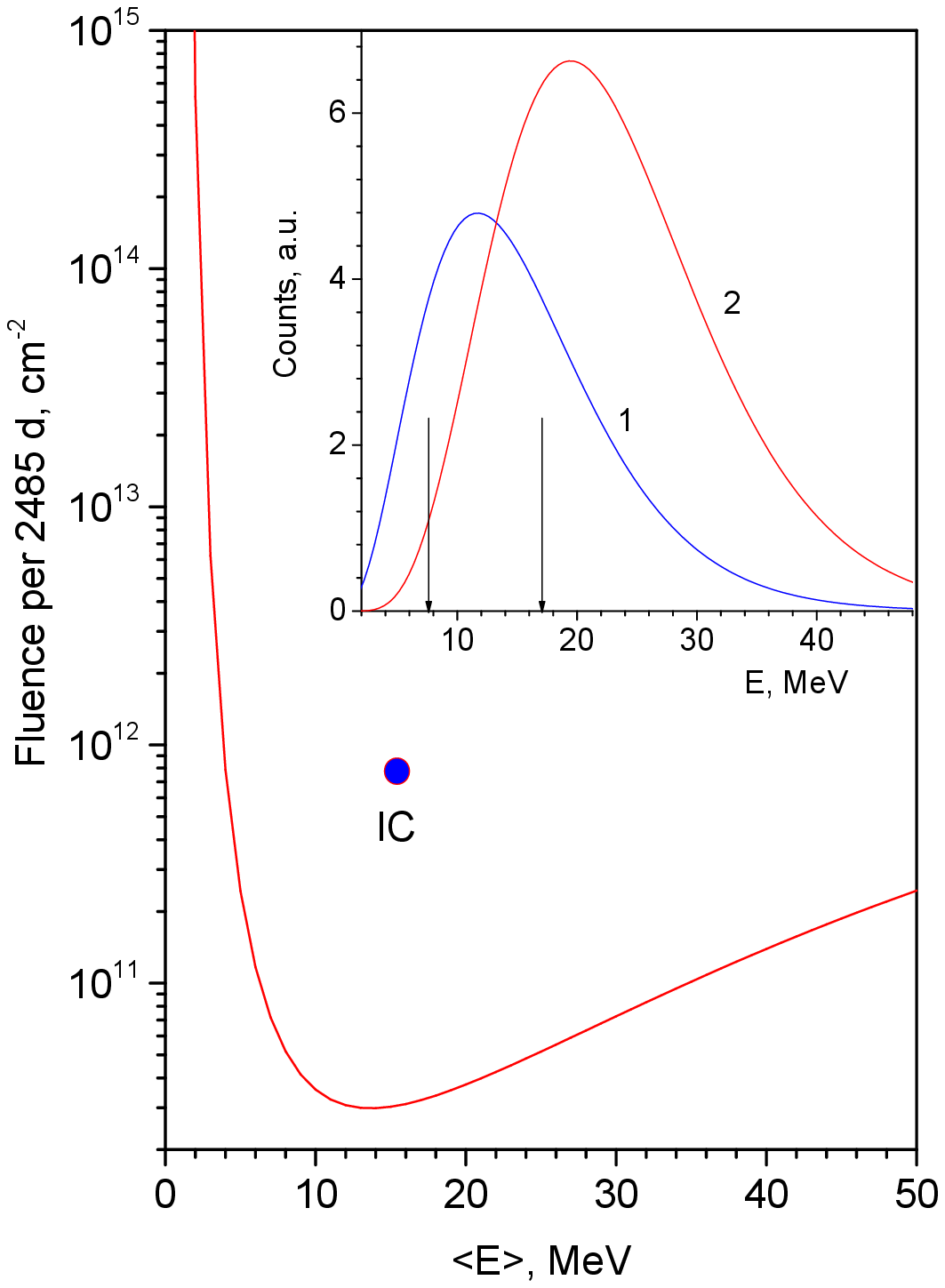}
	\caption{90\%~C.L. upper limits on the $\bar{\nu}_e$ fluence obtained with the SN neutrino spectrum with respect to the mean neutrino energy $\langle E \rangle$. The bold point shows the IceCube result~\cite{Aar2020}. The inset shows the neutrino spectrum for $\langle E \rangle = 15.6$~MeV~(1) and expected Borexino spectrum~(2). Arrows mark the analysis interval.}\label{fig:10}
\end{figure}

In order to estimate a limit on the $\bar{\nu}_e$ fluence, we also exploited the fact that no events were observed with a prompt visible energy exceeding $7.8$~MeV. Assuming two different neutrino energy spectra of monoenergetic and supernova neutrinos and no observed events within the $7.8 -16.8$~MeV interval, we obtain the upper limits on the $\bar{\nu}_e$ fluences presented in Figure~\ref{fig:9}, Figure~\ref{fig:10}, Table \ref{tab:2} and Table  \ref{tab:3} (sixth column in each Table). 

The limits on the fluences of monoenergetic $\bar{\nu}_e$ with the energies below $7.8$~MeV are based on the data from~\cite{Ago2021} in which the upper limits on the $\bar{\nu}_e$ flux in the $1.8 - 7.8$~MeV range were established for $1$~MeV bins. 
The resulting conservative limits obtained by the Feldman-Cousins procedure with the expected background $n_{\rm{bkg}}$ (excluding the cosmogenic component \cite{Ago2021}) are shown in Figure~\ref{fig:9} and Table~\ref{tab:2}.

The limits in Figures~\ref{fig:9} and~\ref{fig:10} are the limits on the total antineutrino fluence over a period of $2485$~days.
Since the expected number of FRBs during this time is $\sim5\times10^6$, the reduced limits per single FRB will be $5\times10^6$ times stronger.

The fluence upper limits can be converted into upper limits on the total energy radiated in the form of neutrinos. 
Here, we consider only the energy radiated by monoenergetic 10 MeV electron (anti)neutrinos assuming an isotropic angular distribution. 
The upper limits on the fluence of $\nu_e$ ($(\nu_e,e)$-scattering) and $\bar{\nu}_e$ (IBD reaction) from the closest flare FRB 200428 lead to restrictions $E\leq 1.3\times10^{53}$~ erg and $E\leq 4.8\times10^{51}$~ erg, correspondingly. 

The average of the inverse square of the distance to the registered FRBs corresponds to $R=400$ Mpc. 
The upper limits for the total FRB fluences of 10 MeV $\nu_e$ and $\bar{\nu}_e$ obtained from spectral fit through the study of $(\nu_e,e)$ elastic scattering and IBD reaction (Table 2) gives weaker limits on the radiated energy $E\leq 4.7\times10^{57}$~ erg/FRB and $E\leq 1.7\times10^{54}$~erg/FRB, correspondingly. This values can be compared with the energy of solar mass $1.8\times 10^{54}$~erg. The most stringent restriction on the release of energy in the form of neutrinos obtained by us for FRB 200428 corresponds to $2.7\times 10^{-3}$ solar mass. 
Limits on the energy radiated into neutrinos of other flavors can be easily calculated from Tables 1, 2 and 3.

Figure~\ref{fig:10} shows the IceCube upper limit on the $\bar{\nu}_e$ fluence of the supernova spectrum with the mean neutrino energy $\langle E \rangle = 15.6$~MeV and pinching the parameter $\alpha$=3 (shown with a circle) based on a collective increase in the rate of hits in the detector in coincidence with 28~FRBs~\cite{Aar2020}.
The IceCube limit can also be compared with the Borexino limit obtained from the temporal correlation analysis (Table 1, line 6, column 6). Due to different selection approach there are only 4 FRBs in the overlap between the sample of 28 FRBs considered in the IceCube analysis~\cite{Aar2020} and our dataset of 42 FRBs with fluence exceeding 40~$\rm{(Jy~ms)}$.
The Borexino detector is capable of detecting single $\bar{\nu}_e$ practically from the threshold of the IBD reaction that leads to significantly higher sensitivity to low energy electron antineutrino fluxes.


\section{Conclusion}

We looked for an excess in the number of events detected by Borexino produced by neutrino-electron elastic scattering and the inverse beta-decay on protons correlated to the most intense FRBs between 2007 and 2021.
We found no statistically significant increase in the number of events, with the visible energy above $0.25$~MeV within time windows of $\pm1000$~s centered at the time of FRB arrivals. 
As a result, new limits on the fluence of monochromatic neutrinos of all flavors were set for neutrino energies in the range of $0.5 - 15$~MeV. 

Another approach was based on the search for specific shapes of neutrino-electron scattering in the high statistic Borexino spectrum.
The strongest limits on the fluences of monoenergetic neutrinos with energies in the range of $0.5 - 50$~MeV and of the supernova neutrino spectrum given by the modified Fermi-Dirac distribution were obtained for different effective neutrino temperatures.
Additionally, the inverse beta-decay reaction was considered to set a new limit on the fluence of electron antineutrinos related to FRBs.  

\begin{acknowledgements}
The Borexino program is possible by funding from INFN (Italy), NSF  (USA), DFG and HGF (Germany), RFBR (Grant 19-02-00097 A), RSF (Grant 21-12-00063) (Russia), and NCN (Grant No. UMO 2013/10/E/ST2/00180) (Poland). We acknowledge the generous hospitality and support of the Laboratori Nazionale del Gran Sasso (Italy).
\end{acknowledgements}



\end{document}